\documentclass[11pt]{article}

\begin{document}
%\begin{document}
%\special{papersize=8.26in,11.69in}
%textwidth17.0cm
%\textheight22.0cm
%\baselineskip1.0cm
%\setlength{\topmargin}{-1cm}
%\addtolength{\textheight}{1cm}
%\oddsidemargin+0.8cm
%\evensidemargin-0.8cm
\pagestyle{plain}
\def\eqa{\!\!&=&\!\!}
\def\ccr{\nonumber\\}

\def\la{\langle}
\def\ra{\rangle}

\def\del{\Delta}
\def\ddel{{}^\bullet\! \Delta}
\def\deld{\Delta^{\hskip -.5mm \bullet}}
\def\ddeld{{}^{\bullet}\! \Delta^{\hskip -.5mm \bullet}}
\def\dddel{{}^{\bullet \bullet} \! \Delta}

\def\rld{\rlap{\,/}D}
\def\rldd{\rlap{\,/}\nabla}
%%%%%%%%%%%%%%% insert actual file mydefs.sty %%%%%%%%%%%%%%%%%%%%%%
%------------------------------------------------------------------------     
% MATH SYMBOLS
%
%fractions
\def\half{{1\over 2}}
\def\third{{1\over3}}
\def\fourth{{1\over4}}
\def\fifth{{1\over5}}
\def\sixth{{1\over6}}
\def\seventh{{1\over7}}
\def\eigth{{1\over8}}
\def\ninth{{1\over9}}
\def\tenth{{1\over10}}
\def\bN{\mathop{\bf N}}
\def\R{{\rm I\!R}}
\def\Eins{{\mathchoice {\rm 1\mskip-4mu l} {\rm 1\mskip-4mu l}
{\rm 1\mskip-4.5mu l} {\rm 1\mskip-5mu l}}}
\def\Z{{\mathchoice {\hbox{$\sf\textstyle Z\kern-0.4em Z$}}
{\hbox{$\sf\textstyle Z\kern-0.4em Z$}}
{\hbox{$\sf\scriptstyle Z\kern-0.3em Z$}}
{\hbox{$\sf\scriptscriptstyle Z\kern-0.2em Z$}}}}
\def\abs#1{\left| #1\right|}
\def\com#1#2{
        \left[#1, #2\right]}
\def\square{\kern1pt\vbox{\hrule height 1.2pt\hbox{\vrule width 1.2pt
   \hskip 3pt\vbox{\vskip 6pt}\hskip 3pt\vrule width 0.6pt}
   \hrule height 0.6pt}\kern1pt}
      \def\boxop{{\raise-.25ex\hbox{\square}}}
% \contract is a differential geometry contraction sign _|
\def\contract{\makebox[1.2em][c]{
        \mbox{\rule{.6em}{.01truein}\rule{.01truein}{.6em}}}}
\def\ltap{\ \raisebox{-.4ex}{\rlap{$\sim$}} \raisebox{.4ex}{$<$}\ }
\def\gtap{\ \raisebox{-.4ex}{\rlap{$\sim$}} \raisebox{.4ex}{$>$}\ }
\def\mn{{\mu\nu}}
\def\rs{{\rho\sigma}}
\newcommand{\Det}{{\rm Det}}
\def\Tr{{\rm Tr}\,}
\def\tr{{\rm tr}\,}
\def\sumij{\sum_{i<j}}
\def\e{\,{\rm e}}
%derivatives
\def\pa{\partial}
\def\dA{\partial^2}
\def\ddx{{d\over dx}}
\def\ddt{{d\over dt}}
\def\der#1#2{{d #1\over d#2}}
\def\lie{\hbox{\it \$}} % fancy L for the Lie derivative
\def\partder#1#2{{\partial #1\over\partial #2}}
\def\secder#1#2#3{{\partial^2 #1\over\partial #2 \partial #3}}
%
%equations
\newcommand{\be}{\begin{equation}}
\newcommand{\ee}{\end{equation}\noindent}
\newcommand{\bear}{\begin{eqnarray}}
\newcommand{\ear}{\end{eqnarray}\noindent}
\newcommand{\benn}{\begin{enumerate}}
\newcommand{\enn}{\end{enumerate}}
\newcommand{\veject}{\vfill\eject}
\newcommand{\ven}{\vfill\eject\noindent}
%
%reference to equations
\def\eq#1{{eq. (\ref{#1})}}
\def\eqs#1#2{{eqs. (\ref{#1}) -- (\ref{#2})}}
%
%integrals
\def\totint{\int_{-\infty}^{\infty}}
\def\posint{\int_0^{\infty}}
\def\negint{\int_{-\infty}^0}
\def\pint{{\dps\int}{dp_i\over {(2\pi)}^d}}
%
% PHYS SYMBOLS
\newcommand{\GeV}{\mbox{GeV}}
\def\FFdual{F\cdot\tilde F}
\def\bra#1{\langle #1 |}
\def\ket#1{| #1 \rangle}
\def\braket#1#2{\langle {#1} \mid {#2} \rangle}
\def\vev#1{\langle #1 \rangle}
\def\rightvac{\mid 0\rangle}
\def\leftvac{\langle 0\mid}
\def\ihbar{{i\over\hbar}}
% dirac matrix stuff
\def\ge{\hbox{$\gamma_1$}}
\def\gz{\hbox{$\gamma_2$}}
\def\gd{\hbox{$\gamma_3$}}
\def\go{\hbox{$\gamma_1$}}
\def\gt{\hbox{\$\gamma_2$}}
\def\gth{\hbox{$\gamma_3$}} 
\def\gf{\hbox{$\gamma_5\;$}}
\def\slash#1{#1\!\!\!\raise.15ex\hbox {/}}
\newcommand{\slD}{\,\raise.15ex\hbox{$/$}\kern-.27em\hbox{$\!\!\!D$}}
\newcommand{\slpartial}{\raise.15ex\hbox{$/$}\kern-.57em\hbox{$\partial$}}
\newcommand{\cL}{\cal L}
\newcommand{\D}{\cal D}
\newcommand{\Dhalf}{{D\over 2}}
\def\eps{\epsilon}
\def\epshalf{{\epsilon\over 2}}
\def\lag{( -\partial^2 + V)}
%worldline
\def\freeexp{{\rm e}^{-\int_0^Td\tau {1\over 4}\dot x^2}}
\def\kinb{{1\over 4}\dot x^2}
\def\kinf{{1\over 2}\psi\dot\psi}
\def\expk{{\rm exp}\biggl[\,\sum_{i<j=1}^4 G_{Bij}k_i\cdot k_j\biggr]}
\def\expp{{\rm exp}\biggl[\,\sum_{i<j=1}^4 G_{Bij}p_i\cdot p_j\biggr]}
\def\expshort{{\e}^{\half G_{Bij}k_i\cdot k_j}}
\def\expabb{{\e}^{(\cdot )}}
\def\epseps#1#2{\varepsilon_{#1}\cdot \varepsilon_{#2}}
\def\epsk#1#2{\varepsilon_{#1}\cdot k_{#2}}
\def\kk#1#2{k_{#1}\cdot k_{#2}}
\def\G#1#2{G_{B#1#2}}
\def\Gp#1#2{{\dot G_{B#1#2}}}
\def\GF#1#2{G_{F#1#2}}
\def\Dab{{(x_a-x_b)}}
\def\Dsq{{({(x_a-x_b)}^2)}}
\def\PITD{{(4\pi T)}^{-{D\over 2}}}
\def\4piTD{{(4\pi T)}^{-{D\over 2}}}
\def\4piT4{{(4\pi T)}^{-2}}
\def\TintmD{{\dps\int_{0}^{\infty}}{dT\over T}\,e^{-m^2T}
    {(4\pi T)}^{-{D\over 2}}}
\def\Tintm4{{\dps\int_{0}^{\infty}}{dT\over T}\,e^{-m^2T}
    {(4\pi T)}^{-2}}
\def\Tintm{{\dps\int_{0}^{\infty}}{dT\over T}\,e^{-m^2T}}
\def\Tint{{\dps\int_{0}^{\infty}}{dT\over T}}
\def\np{n_{+}}
\def\nm{n_{-}}
\def\Np{N_{+}}
\def\Nm{N_{-}}
\newcommand{\slG}{{{\dot G}\!\!\!\! \raise.15ex\hbox {/}}}
\newcommand{\Gd}{{\dot G}}
\newcommand{\Gund}{{\underline{\dot G}}}
\newcommand{\Gdd}{{\ddot G}}
\def\GBd12{{\dot G}_{B12}}
\def\Dx{\dps\int{\cal D}x}
\def\Dy{\dps\int{\cal D}y}
\def\Dpsi{\dps\int{\cal D}\psi}
\def\dint#1{\int\!\!\!\!\!\int\limits_{\!\!#1}}
\def\ddtau{{d\over d\tau}}
\def\ie{\hbox{$\textstyle{\int_1}$}}
\def\iz{\hbox{$\textstyle{\int_2}$}}
\def\id{\hbox{$\textstyle{\int_3}$}}
\def\ldop{\hbox{$\lbrace\mskip -4.5mu\mid$}}
\def\rdop{\hbox{$\mid\mskip -4.3mu\rbrace$}}
%
%VARIOUS
\newcommand{\1}{{\'\i}}
\newcommand{\no}{\noindent}
\def\non{\nonumber}
\def\dps{\displaystyle}
\def\sy{\scriptscriptstyle}
\def\sy{\scriptscriptstyle}
%-------------------------------------------------------------------------

%-------------------------------------------------------------------------
%% My own definitions
% From cv.sty
%
% Title page

\newcommand{\bea}{\begin{eqnarray}}  
\newcommand{\eea}{\end{eqnarray}}  
\def\eqa{&=&}  
\def\ccr{\nonumber\\}  
  
\def\a{\alpha}
\def\b{\beta}
\def\m{\mu}
\def\n{\nu}
\def\r{\rho}
\def\s{\sigma}
\def\ep{\epsilon}

\def\cosech{\rm cosech}
\def\sech{\rm sech}
\def\coth{\rm coth}
\def\tanh{\rm tanh}

%%%%%%%%  
\def\sqr#1#2{{\vcenter{\vbox{\hrule height.#2pt  
     \hbox{\vrule width.#2pt height#1pt \kern#1pt  
           \vrule width.#2pt}  
       \hrule height.#2pt}}}}  
\def\square{\mathchoice\sqr66\sqr66\sqr{2.1}3\sqr{1.5}3}  
%%%%%%%%%%%  
  
\def\appendix{\par\clearpage
  \setcounter{section}{0}
  \setcounter{subsection}{0}
 % \@addtoreset{equation}{section}
  \def\@sectname{Appendix~}
  \def\theequation{\thesection\arabic{equation}}
  \def\thesection{\Alph{section}}}
 
% Figures
\def\thefigures#1{\par\clearpage\section*{Figures\@mkboth
  {FIGURES}{FIGURES}}\list
  {Fig.~\arabic{enumi}.}{\labelwidth\parindent\advance
\labelwidth -\labelsep
      \leftmargin\parindent\usecounter{enumi}}}
\def\figitem#1{\item\label{#1}}
\let\endthefigures=\endlist
 
% Tables
\def\thetables#1{\par\clearpage\section*{Tables\@mkboth
  {TABLES}{TABLES}}\list
  {Table~\Roman{enumi}.}{\labelwidth-\labelsep
      \leftmargin0pt\usecounter{enumi}}}
\def\tableitem#1{\item\label{#1}}
\let\endthetables=\endlist
 
% Put period after section number and allow for APPENDIX prefix.
\def\@sect#1#2#3#4#5#6[#7]#8{\ifnum #2>\c@secnumdepth
     \def\@svsec{}\else
     \refstepcounter{#1}\edef\@svsec{\@sectname\csname the#1\endcsname
.\hskip 1em }\fi
     \@tempskipa #5\relax
      \ifdim \@tempskipa>\z@
        \begingroup #6\relax
          \@hangfrom{\hskip #3\relax\@svsec}{\interlinepenalty \@M #8\par}
        \endgroup
       \csname #1mark\endcsname{#7}\addcontentsline
         {toc}{#1}{\ifnum #2>\c@secnumdepth \else
                      \protect\numberline{\csname the#1\endcsname}\fi
                    #7}\else
        \def\@svse=chd{#6\hskip #3\@svsec #8\csname #1mark\endcsname
                      {#7}\addcontentsline
                           {toc}{#1}{\ifnum #2>\c@secnumdepth \else
                             \protect\numberline{\csname the#1\endcsname}\fi
                       #7}}\fi
     \@xsect{#5}}
 
\def\@sectname{}
%
%                 M A T E X
%
%       This defines et al., i.e., e.g., cf., etc.
\def\eg{\hbox{\it e.g.}}        \def\cf{\hbox{\it cf.}}
\def\etal{\hbox{\it et al.}}
\def\dash{\hbox{---}}
%       common physics symbols
\def\bR{\mathop{\bf R}}
\def\bC{\mathop{\bf C}}
\def\eq#1{{eq. \ref{#1}}}
\def\eqs#1#2{{eqs. \ref{#1}--\ref{#2}}}
\def\lie{\hbox{\it \$}} % fancy L for the Lie derivative
\def\partder#1#2{{\partial #1\over\partial #2}}
\def\secder#1#2#3{{\partial^2 #1\over\partial #2 \partial #3}}
\def\abs#1{\left| #1\right|}
\def\ltap{\ \raisebox{-.4ex}{\rlap{$\sim$}} \raisebox{.4ex}{$<$}\ }
\def\gtap{\ \raisebox{-.4ex}{\rlap{$\sim$}} \raisebox{.4ex}{$>$}\ }
% \contract is a differential geometry contraction sign _|
\def\contract{\makebox[1.2em][c]{
        \mbox{\rule{.6em}{.01truein}\rule{.01truein}{.6em}}}}
% double-headed superior arrow added 9.2.86
%
% commutator added 11.14.86
\def\com#1#2{
        \left[#1, #2\right]}
%
%these written by orlando alvarez
% ************************************************************
%       The following macros were written by Chris Quigg.
%       They create bent arrows and can be used to write
%       decays such as pi --> mu + nu
%                              --> e nu nubar
%
\def\bentarrow{\:\raisebox{1.3ex}{\rlap{$\vert$}}\!\rightarrow}
\def\longbent{\:\raisebox{3.5ex}{\rlap{$\vert$}}\raisebox{1.3ex}%
        {\rlap{$\vert$}}\!\rightarrow}
\def\onedk#1#2{
        \begin{equation}
        \begin{array}{l}
         #1 \\
         \bentarrow #2
        \end{array}
        \end{equation}
                }
\def\dk#1#2#3{
        \begin{equation}
        \begin{array}{r c l}
        #1 & \rightarrow & #2 \\
         & & \bentarrow #3
        \end{array}
        \end{equation}
                }
\def\dkp#1#2#3#4{
        \begin{equation}
        \begin{array}{r c l}
        #1 & \rightarrow & #2#3 \\
         & & \phantom{\; #2}\bentarrow #4
        \end{array}
        \end{equation}
                }
\def\bothdk#1#2#3#4#5{
        \begin{equation}
        \begin{array}{r c l}
        #1 & \rightarrow & #2#3 \\
         & & \:\raisebox{1.3ex}{\rlap{$\vert$}}\raisebox{-0.5ex}{$\vert$}%
        \phantom{#2}\!\bentarrow #4 \\
         & & \bentarrow #5
        \end{array}
        \end{equation}
                }
AEI-2009-081
\begin{center}
{\Large\bf Effective action for Einstein-Maxwell theory}\\
{\Large\bf at order $RF^4$}\\

\bigskip

{Jos\'e Manuel D\'avila$^{b}$, 
Christian Schubert$^{a,b}$}
\begin{itemize}
\item [$^a$]
{\it 
%   \footnote{}
Max-Planck-Institut f\"ur Gravitationsphysik, Albert-Einstein-Institut,
M\"uhlenberg 1, D-14476 Potsdam, Germany
}
\item [$^b$]
{\it 
Instituto de F\'{\i}sica y Matem\'aticas
\\
Universidad Michoacana de San Nicol\'as de Hidalgo\\
Edificio C-3, Apdo. Postal 2-82\\
C.P. 58040, Morelia, Michoac\'an, M\'exico\\
}
\end{itemize}
\end{center}

\bigskip

\noindent
{\bf Abstract:}
We use a recently derived integral representation of the 
one-loop effective action in Einstein-Maxwell theory
for an explicit calculation of the part of the effective action
containing the information on the low energy limit of the five-point amplitudes
involving one graviton, four photons and either a scalar or
spinor loop. All available identities are used to get the result
into a relatively compact form.

\vfill\eject\noindent

\renewcommand{\theequation}{\arabic{section}.\arabic{equation}}
\renewcommand{\arraystretch}{2.5}
\def\R{1\!\!{\rm R}}
\def\Eins{\mathord{1\hskip -1.5pt
\vrule width .5pt height 7.75pt depth -.2pt \hskip -1.2pt
\vrule width 2.5pt height .3pt depth -.05pt \hskip 1.5pt}}
\newcommand{\symb}{\mbox{symb}}
\renewcommand{\arraystretch}{2.5}
\def\GBd12{{\dot G}_{B12}}
\def\mneg{\!\!\!\!\!\!\!\!\!\!}
\def\Mneg{\!\!\!\!\!\!\!\!\!\!\!\!\!\!\!\!\!\!\!\!}
\def\non{\nonumber}
\def\beqn*{\begin{eqnarray*}}
\def\eqn*{\end{eqnarray*}}
\def\sy{\scriptscriptstyle}
\def\footstrut{\baselineskip 12pt}
\def\square{\kern1pt\vbox{\hrule height 1.2pt\hbox{\vrule width 1.2pt
   \hskip 3pt\vbox{\vskip 6pt}\hskip 3pt\vrule width 0.6pt}
   \hrule height 0.6pt}\kern1pt}
\def\np{n_{+}}
\def\nm{n_{-}}
\def\Np{N_{+}}
\def\Nm{N_{-}}
\def\exmn{\Bigl(\mu \leftrightarrow \nu \Bigr)}
\def\slash#1{#1\!\!\!\raise.15ex\hbox {/}}
\def\dint#1{\int\!\!\!\!\!\int\limits_{\!\!#1}}
\def\bra#1{\langle #1 |}
\def\ket#1{| #1 \rangle}
\def\vev#1{\langle #1 \rangle}
\def\rightvac{\mid 0\rangle}
\def\leftvac{\langle 0\mid}
\def\dps{\displaystyle}
\def\sy{\scriptscriptstyle}
\def\half{{1\over 2}}
\def\third{{1\over3}}
\def\fourth{{1\over4}}
\def\fifth{{1\over5}}
\def\sixth{{1\over6}}
\def\seventh{{1\over7}}
\def\eigth{{1\over8}}
\def\ninth{{1\over9}}
\def\tenth{{1\over10}}
\def\pa{\partial}
\def\ddtau{{d\over d\tau}}
\def\ge{\hbox{\textfont1=\tame $\gamma_1$}}
\def\gz{\hbox{\textfont1=\tame $\gamma_2$}}
\def\gd{\hbox{\textfont1=\tame $\gamma_3$}}
\def\go{\hbox{\textfont1=\tamt $\gamma_1$}}
\def\gt{\hbox{\textfont1=\tamt $\gamma_2$}}
\def\gth{\hbox{\textfont1=\tamt $\gamma_3$}} 
\def\gf{\hbox{$\gamma_5\;$}}
\def\ie{\hbox{$\textstyle{\int_1}$}}
\def\iz{\hbox{$\textstyle{\int_2}$}}
\def\id{\hbox{$\textstyle{\int_3}$}}
\def\ldop{\hbox{$\lbrace\mskip -4.5mu\mid$}}
\def\rdop{\hbox{$\mid\mskip -4.3mu\rbrace$}}
\def\eps{\epsilon}
\def\epshalf{{\epsilon\over 2}}
\def\e{\mbox{e}}
\def\mn{{\mu\nu}}
\def\exmn{{(\mu\leftrightarrow\nu )}}
\def\ab{{\alpha\beta}}
\def\exab{{(\alpha\leftrightarrow\beta )}}
\def\g{\mbox{g}}
\def\kinb{{1\over 4}\dot x^2}
\def\kinf{{1\over 2}\psi\dot\psi}
\def\expk{{\rm exp}\biggl[\,\sum_{i<j=1}^4 G_{Bij}k_i\cdot k_j\biggr]}
\def\expp{{\rm exp}\biggl[\,\sum_{i<j=1}^4 G_{Bij}p_i\cdot p_j\biggr]}
\def\expshort{{\e}^{\half G_{Bij}k_i\cdot k_j}}
\def\expabb{{\e}^{(\cdot )}}
\def\epseps#1#2{\varepsilon_{#1}\cdot \varepsilon_{#2}}
\def\epsk#1#2{\varepsilon_{#1}\cdot k_{#2}}
\def\kk#1#2{k_{#1}\cdot k_{#2}}
\def\G#1#2{G_{B#1#2}}
\def\Gp#1#2{{\dot G_{B#1#2}}}
\def\GF#1#2{G_{F#1#2}}
\def\Dab{{(x_a-x_b)}}
\def\Dsq{{({(x_a-x_b)}^2)}}
\def\lag{( -\partial^2 + V)}
\def\PITD{{(4\pi T)}^{-{D\over 2}}}
\def\4piTD{{(4\pi T)}^{-{D\over 2}}}
\def\4piT4{{(4\pi T)}^{-2}}
\def\TintmD{{\dps\int_{0}^{\infty}}{dT\over T}\,e^{-m^2T}
    {(4\pi T)}^{-{D\over 2}}}
\def\Tintm4{{\dps\int_{0}^{\infty}}{dT\over T}\,e^{-m^2T}
    {(4\pi T)}^{-2}}
\def\Tintm{{\dps\int_{0}^{\infty}}{dT\over T}\,e^{-m^2T}}
\def\Tint{{\dps\int_{0}^{\infty}}{dT\over T}}
\def\pint{{\dps\int}{dp_i\over {(2\pi)}^d}}
\def\Dx{\dps\int{\cal D}x}
\def\Dy{\dps\int{\cal D}y}
\def\Dpsi{\dps\int{\cal D}\psi}
\def\Tr{{\rm Tr}\,}
\def\tr{{\rm tr}\,}
\def\sumij{\sum_{i<j}}
\def\freeexp{{\rm e}^{-\int_0^Td\tau {1\over 4}\dot x^2}}
\def\arraystretch{2.5}
\def\Ge{\mbox{GeV}}
\def\dA{\partial^2}
\def\DA{\sqsubset\!\!\!\!\sqsupset}
\def\FFdual{F\cdot\tilde F}
\def\mn{{\mu\nu}}
\def\rs{{\rho\sigma}}
\def\oplusotimes{{{\lower 15pt\hbox{$\scriptscriptstyle \oplus$}}\atop{\otimes}}}
\def\perppar{{{\lower 15pt\hbox{$\scriptscriptstyle \perp$}}\atop{\parallel}}}
\def\oopp{{{\lower 15pt\hbox{$\scriptscriptstyle \oplus$}}\atop{\otimes}}\!{{\lower 15pt\hbox{$\scriptscriptstyle \perp$}}\atop{\parallel}}}
%\font\tame = cmmi12 scaled\magstep1
%\font\tamt = cmmi12 scaled\magstep2
%-------------------------------------------------------------------------
% To change the LaTeX pagestyle
% example  DINA4 format DESY
%\newlength{\dinwidth}
%\newlength{\dinmargin}
%\setlength{\dinwidth}{21.0cm}
%\textheight23.2cm
%\textwidth17.0cm
%\setlength{\dinmargin}{\dinwidth}
%\addtolength{\dinmargin}{-\textwidth}
%\setlength{\dinmargin}{0.5\dinmargin}
%\oddsidemargin -1.0in
%\addtolength{\oddsidemargin}{\dinmargin}
%\setlength{\evensidemargin}{\oddsidemargin}
%\setlength{\marginparwidth}{0.9\dinmargin}
%\marginparsep 8pt \marginparpush 5pt
%\topmargin -42pt
%\headheight 12pt 
%\headsep 30pt \footheight 12pt \footskip
%24pt
%-----------------------------------------------------------------------
% uncomment any of these if you want numbering to respect the sections
%
% \renewcommand{\thesection}{\arabic{section}.}
% \renewcommand{\thesubsection}{\thesection\arabic{subsection}.}
% \renewcommand{\theequation}{{\protect\thesection\arabic{equation}}}
% \renewcommand{\thetable}{{\protect{\bf \thesection\arabic{table}}}}
% \renewcommand{\thetable}{{\protect{\thesection\arabic{table}}}}
% \renewcommand{\thefigure}{{\protect\bf\thesection\arabic{figure}}}
% \renewcommand{\thefigure}{{\protect\thesection\arabic{figure}}}
% \renewcommand{\textfraction}{0}
% \renewcommand{\topfraction}{1.00}
% \renewcommand{\bottomfraction}{1.00}
% \renewcommand{\baselinestretch}{1.1}
%-----------------------------------------------------------------------
% special symbols: real numbers, unit matrix, integers
%
\def\bbbr{{\rm I\!R}}
\def\bbbone{{\mathchoice {\rm 1\mskip-4mu l} {\rm 1\mskip-4mu l}
{\rm 1\mskip-4.5mu l} {\rm 1\mskip-5mu l}}}
\def\bbbz{{\mathchoice {\hbox{$\sf\textstyle Z\kern-0.4em Z$}}
{\hbox{$\sf\textstyle Z\kern-0.4em Z$}}
{\hbox{$\sf\scriptstyle Z\kern-0.3em Z$}}
{\hbox{$\sf\scriptscriptstyle Z\kern-0.2em Z$}}}}

%%%%%%%%%%%%%%%%%%%%%%%%%%%%%%%%%%%%%%%%%%%%%%%%%%%%%%%%%%%%%%%%%%%%%%%
\renewcommand{\thefootnote}{\protect\arabic{footnote}}
%\pagestyle{plain}
%------------------------------------------------------
%\hfill {\large AEI-2008-053}
%
%\hfill {\large UMSNH-IFM-F-2008-24}

%\pagestyle{plain}
%\setcounter{page}{1}
%\setcounter{footnote}{0}

%\topmargin=-.3in
%\textheight=8.3in
%\textwidth=7.2in
%\textwidth=6.7in

\vspace{10pt}
\section{Introduction}
\renewcommand{\theequation}{1.\arabic{equation}}
\setcounter{equation}{0}

In recent years, much effort has been devoted to the study of the
structure of graviton amplitudes. This was largely due to
developments in string theory, which led to the prediction
that such amplitudes should be much more closely related to
gauge theory amplitudes than one would suspect by comparing 
the Lagrangians or Feynman rules of gravitational and gauge
theories. Specifically, the Kawai-Lewellen-Tye (KLT) 
relations in string theory imply that graviton amplitudes
should be ``squares'' of gauge theory amplitudes \cite{kalety,begiku,bdpr,bern,anathe}.
String theory was also instrumental in providing guiding
principles to develop new powerful techniques for the computation
of graviton amplitudes \cite{bedush,bbst,cacsvr,beboca}.  
Additional motivation comes from the possible
finiteness of $N=8$ supergravity (see \cite{bcdjr} and refs. therein).

This work was largely confined to the case of massless on-shell amplitudes,
for which particularly efficient computation methods are available.
Relatively little seems to have been done on amplitudes involving the
interaction of gravitons with massive matter. At tree level, there are some
classic results on amplitudes involving gravitons \cite{bergas,milton}.
More recently, the tree-level Compton-type amplitudes involving gravitons and
spin zero, half and one particles were computed \cite{holstein} to verify another remarkable
factorization property \cite{chshso} of the graviton-graviton scattering amplitudes in terms
of the photonic Compton amplitudes. 

However, we are not aware of results on graviton amplitudes involving a massive loop, other than
the cases of the graviton propagator \cite{capper,baszir} and of photon-graviton
conversion \cite{phograv1,phograv2}. 
We believe that new insight into the structural relations between photon and graviton
amplitudes might be obtained by 
studying the $N$ graviton amplitudes involving a massive loop,
and more generally the mixed one-loop graviton-photon amplitudes.
Generally, massive one-loop $N$ - point amplitudes are significantly more
difficult to compute than massless ones; on the other hand, their large mass
limit is quite accessible through the effective action. 
For the prototypical case, the QED $N$ - photon amplitude, the information
on the large mass limit is contained in the Euler-Heisenberg
Lagrangian (``EHL'') \cite{eulhei}. We recall the standard proper time representation
of this effective Lagrangian:

\bear
{\cal L}_{\rm spin} &=& - {1\over 8\pi^2}
\int_0^{\infty}{dT\over T^3}
\,\e^{-m^2T}
\biggl\lbrack
{(eaT)(ebT)\over {\rm tanh}(eaT)\tan(ebT)} 
%\nonumber\\&&\hspace{70pt}
- {e^2\over 3}(a^2-b^2)T^2 -1
\biggr\rbrack \, .
\non\\
\label{ehspin}
\ear
Here $T$ is the proper-time of the loop fermion, $m$ its mass, and $a,b$ are the two
Maxwell field invariants,
related to $\bf E$, $\bf B$ by $a^2-b^2 = B^2-E^2,\quad ab = {\bf E}\cdot {\bf B}$.
%The subtraction terms implement the renormalization of charge and vacuum
%energy.
The analogous representation for scalar QED is due to Weisskopf \cite{weisskopf}.

After expanding the  EHL in powers of the field invariants, it is straightforward to
obtain the large mass limit of the $N$ photon amplitudes from 
the terms in this expansion involving $N$ powers of the field. This limit is, of course,
also the limit of low photon energies. The result of this procedure can be
expressed quite concisely \cite{mascvi}:

\bear
\Gamma_{\rm spin}^{(EH)}
[\varepsilon_1^+;\ldots ;\varepsilon_K^+;\varepsilon_{K+1}^-;\ldots ;\varepsilon_N^-]
&=&
-{m^4\over 8\pi^2}
\Bigl({2ie\over m^2}\Bigr)^N(N-3)!
\non\\&&\hspace{-90pt}\times
\sum_{k=0}^K\sum_{l=0}^{N-K}
(-1)^{N-K-l}
{{\cal B}_{k+l}{\cal B}_{N-k-l}
\over
k!l!(K-k)!(N-K-l)!}
\chi_K^+\chi_{N-K}^- 
\, .
\non\\
\label{resspin}
\ear
Here the superscripts $\pm$ refer to circular polarizations, and the ${\cal B}_k$ are
Bernoulli numbers. The invariants $\chi_K^{\pm}$ are written, in standard spinor helicity
notation,

\bear
\chi_K^+ &=&
{({\frac{K}{2}})!
\over 2^{K\over 2}}
\Bigl\lbrace
[12]^2[34]^2\cdots [(K-1)K]^2 + {\rm \,\, all \,\, permutations}
\Bigr\rbrace,
\non\\
\chi_{N-K}^- &=&
{({\frac{N-K}{2}})!
\over 2^{N-K\over 2}}
\Bigl\lbrace
\langle (K+1)(K+2)\rangle^2\langle (K+3)(K+4)\rangle^2\cdots
\langle (N-1)N\rangle^2 + {\rm \,\, all \,\, perm.}
\Bigr\rbrace.
\non\\
\label{defchiKL+-}
\ear 
A very similar formula results for the scalar loop case \cite{mascvi}. 
For the case of the "maximally helicity-violating" (MHV) amplitudes, which have
all $"+"$ or all $"-"$ helicities, eq. (\ref{resspin}) and its scalar analogue have
been generalized to the two-loop level \cite{dunschsd1}. A recently discovered 
correspondence of effective actions points to a relation between scalar loop MHV 
photon amplitudes in $2n$ dimensions and spinor loop graviton amplitudes in 
$4n$ dimensions \cite{basdun}.

One of the long-term goals of the present line of work is to obtain a generalization of
(\ref{resspin}) to the case of the mixed $N$ - photon / $M$ - graviton amplitudes.
As a first step, in \cite{badasc} the EHL (\ref{ehspin}) and its scalar analogue
were generalized to the case relevant for the case of the amplitudes involving
$N$ photons and just one graviton. This corresponded to calculating the one-loop
effective action in  scalar and spinor Einstein-Maxwell theory, 
to all orders in the electromagnetic field strength, and to leading order in the curvature,
also including terms where the curvature tensor gets replaced by two covariant
derivatives. 
These integral representations are given below in section \ref{oldresults} for easy
reference. Although they contain the full information on the low energy limit of the
$N$ - photon / one graviton amplitudes, it is, contrary to the Euler-Heisenberg case, 
a nontrivial task to expand them out in powers of
the field invariants and extract the explicit form of those amplitudes. 
In \cite{badasc} this was done at the $F^2$ level, as a check of consistency with
previous results in the literature. In particular, the $F^2$ part for the spinor loop
was shown to coincide, up to total derivative terms, with the effective
Lagrangian obtained first by Drummond and Hathrell \cite{druhat},

\bear
{\cal L}_{\rm spin}^{(DH)} &=& 
\frac{1}{180 (4\pi)^2m^2} \bigg(
5 R F_{\mu\nu}^2 
-26 R_{\mu\nu} F^{\mu\alpha} F^\nu{}_\alpha
+2  R_{\mu\nu\alpha\beta}F^{\mu\nu}F^{\alpha\beta} \non\\
&& \qquad\qquad
+24 (\nabla^\alpha F_{\alpha\mu})^2  
\bigg )
\label{drumhath}
\ear
(here and in the following we will absorb the electric charge $e$ into the field strength tensor $F$).

In this note, we present the next order in the expansion of the 
effective Lagrangians obtained in \cite{badasc} in powers of the field strength, 
i.e. the terms of order $RF^4$ (there are no order $RF^3$ terms for parity reasons). 
The explicit form of these Lagrangians is given in section \ref{newresults},
in a form made as compact as possible by the use of the gauge and gravitational
Bianchi identities.

\section{Gravitational Euler-Heisenberg Lagrangians to order $R$}
\label{oldresults}
\renewcommand{\theequation}{2.\arabic{equation}}
\setcounter{equation}{0}

In \cite{badasc} Euler-Heisenberg type integral representations were obtained for the
scalar and spinor loop effective Lagrangians in the approximation discussed above.
For the spinor loop, the result reads

\begin{eqnarray}
{\cal L}_{\rm spin}^R &=&
-{1\over 8\pi^2}
\int^{\infty}_{0} \frac{dT}{T^3}\,\e^{-m^2T}\mbox{det}^{-1/2}\left[ \frac{\tan(FT)}{FT}\right] 
\nonumber\\&&\times
\Biggl\lbrace
1+\frac{iT^2}{8}F_{\mu \nu ;  \alpha \beta}\,\,{\cal G}^{\alpha \beta}_{B11}\Big(\dot{{\cal G}}^{\mu \nu}_{B11}-2\,{\cal G}^{\mu \nu}_{F11} \Big) \nonumber\\
&&+\frac{i T^2}{8}\left(F_{\mu \nu ; \beta \alpha} + F_{\mu \nu ;  \alpha \beta}\right)\dot{{\cal G}}^{\mu \beta}_{B11}{\cal G}^{\nu \alpha}_{B11}+\frac{T}{3}R_{\alpha \beta}\,{\cal G}^{\alpha \beta}_{B11} \nonumber\\
&&-\frac{i T^2}{24}F_{\lambda \nu}R^{\lambda}_{\, \, \, \alpha \beta\mu}\,\left(\dot{{\cal G}}^{\nu \mu}_{B11}\,{\cal G}^{\alpha \beta}_{B11}+\dot{{\cal G}}^{\alpha \mu}_{B11}\,{\cal G}^{\nu \beta}_{B11}+\dot{{\cal G}}^{\beta \mu}_{B11}\,{\cal G}^{\nu \alpha}_{B11}+4\,{\cal G}^{\mu \nu}_{F11}\,{\cal G}^{\alpha \beta}_{B11}\right) \nonumber\\
&&+\frac{T}{12}R_{\mu \alpha \beta \nu}\Big(\dot{{\cal G}}^{\mu \alpha}_{B11}\dot{{\cal G}}^{\beta \nu}_{B11}+\dot{{\cal G}}^{\mu \beta}_{B11}\dot{{\cal G}}^{\alpha\nu}_{B11}
+\Bigl(\ddot{{\cal G}}^{\mu \nu}_{B11}-2g^\mn\delta(0)\Bigr){\cal G}^{\alpha \beta}_{B11}
\non\\
&&+\dot{{\cal G}}^{\alpha \beta}_{B11}\,{\cal G}^{\mu \nu}_{F11}
+\dot{{\cal G}}^{\nu \beta}_{B11}\,{\cal G}^{\mu \alpha}_{F11}
-{\cal G}^{\alpha \beta}_{B11}\,\Bigl(\dot{{\cal G}}^{\mu \nu}_{F11}-2g^\mn\delta(0)\Bigr)
\Big) \nonumber\\
&&-\frac{1}{6}T^{3}F_{\alpha \beta; \gamma}\,F_{\mu \nu ; \delta}\,\int^{1}_{0}d\tau_{1}\Big(\dot{{\cal G}}^{\alpha \nu}_{B12}\,\dot{{\cal G}}^{\beta \mu}_{B12} \, {\cal G}^{\gamma \delta}_{B12}+\dot{{\cal G}}^{\alpha \nu}_{B12}\,{\cal G}^{\beta \delta}_{B12} \, \dot{{\cal G}}^{\gamma \mu}_{B12} \nonumber\\
&&+\frac{3}{2}\,{\cal G}^{\gamma \delta}_{B12}\,{\cal G}^{\alpha \mu}_{F12}\,{\cal G}^{\beta \nu}_{F12}
%- \frac{3}{4}\, {\cal G}^{\gamma \eta}_{B12}{\cal G}^{\alpha \beta}_{F11}\,{\cal G}^{\mu \nu}_{F22}\nonumber\\
%&&+\frac{1}{T^2}{\cal G}^{\mu \nu}_{F22}\dot{{\cal G}}^{\alpha \beta}_{B11}\,{\cal G}^{\gamma \eta}_{B12}+\frac{1}{T^2}{\cal G}^{\mu \nu}_{F22}\dot{{\cal G}}^{\alpha \gamma}_{B11}\,{\cal G}^{\beta \eta}_{B12}+\frac{1}{T^2}{\cal G}^{\mu \nu}_{F22}\dot{{\cal G}}^{\alpha \eta}_{B12}\,{\cal G}^{\beta \gamma}_{B11}        
\Big)  \Biggr\rbrace  .   \nonumber\\
\label{resultspin}		
\end{eqnarray}
Here the determinant factor $\mbox{det}^{-1/2}\left[ \frac{\tan(FT)}{FT}\right]$ by itself
would just reproduce the (unrenormalized) Euler-Heisenberg Lagrangian (\ref{ehspin}).
The integrand involves the worldline Green's functions in a constant field, as well as their
derivatives. Those Green's functions can be written as 

\bear
{\cal G}_{B12} &\equiv& {\cal G}_B (\tau_1,\tau_2) =
{1\over 2{\cal Z}^2}\Biggl({{\cal Z}\over{{\rm sin}({\cal Z})}}
{\rm e}^{-i{\cal Z}\dot G_{B12}}
\!+\! i{\cal Z}\dot G_{B12} - 1\Biggr) \, ,
 \non\\
\dot{\cal G}_{B12} &\equiv& \frac{\partial}{\partial\tau_1}{\cal G}_B(\tau_1,\tau_2)
=
{i\over {\cal Z}}\biggl({{\cal Z}\over{{\rm sin}({\cal Z})}}
\,{\rm e}^{-i{\cal Z}\dot G_{B12}}-1\biggr) \, ,
\nonumber\\
\ddot{\cal G}_{B12} &\equiv & 
\frac{\partial^2}{\partial\tau_1^2}{\cal G}_{B}(\tau_1,\tau_2)
= 2\delta(\tau_1-\tau_2) -2{{\cal Z}\over{{\rm sin}({\cal Z})}}
\,{\rm e}^{-i{\cal Z}\dot G_{B12}} \, ,\nonumber\\
{\cal G}_{F12} &\equiv& {\cal G}_F (\tau_1,\tau_2) =
G_{F12}
{{\rm e}^{-i{\cal Z}\dot G_{B12}}\over \cos ({\cal Z})} \, ,
 \non\\
 \dot{\cal G}_{F12} &\equiv& 
\frac{\partial}{\partial\tau_1} {\cal G}_F(\tau_1,\tau_2)
= 2\delta(\tau_1-\tau_2) + 2iG_{F12}{{\cal Z}\over{{\rm cos}({\cal Z})}}
\, {\rm e}^{-i{\cal Z}\dot G_{B12}} \, ,
\non\\
 \label{calGBGF}
\ear
with $\dot G_{B12}={\rm sign}(\tau_1-\tau_2)-2(\tau_1-\tau_2),  G_{F12}={\rm sign}(\tau_1-\tau_2)$.
The right hand sides of eqs.(\ref{calGBGF}) are to be understood as power series in the matrix 
${\cal Z}_\mn := TF_\mn (x_0)$, where
the indices are raised and lowered with $g_\mn (x_0)$.
We remark that the explicit $\delta(0)$'s in (\ref{resultspin}) subtract other $\delta(0)$'s contained in the coincidence limits $\ddot {\cal G}_{B11}$ and $\dot{\cal G}_{F11}$ \cite{badasc}.
The Green's functions in (\ref{resultspin}) with two different indices (i.e. which are not
coincidence limits) are understood to have $\tau_2 = 0$.

\no
For the case of a scalar in the loop, the result is somewhat simpler:

\begin{eqnarray}
{\cal L}_{\rm scal}^R&=&{1\over 16\pi^2}
\int^{\infty}_{0} \frac{dT}{T^3}\,\e^{-m^2T}
\mbox{det}^{-1/2}\left[ \frac{\sin(FT)}{FT}\right]\Biggl\lbrace 1-T\bar\xi R 
+\frac{T}{3}{\cal G}^{\alpha \beta}_{B11}R_{\alpha \beta}\nonumber\\
&&  +\frac{iT^2}{8}F_{\mu \nu ;  \alpha \beta}\,\dot{{\cal G}}^{\mu \nu}_{B11}\,{\cal G}^{\alpha \beta}_{B11}+\frac{i}{8}T^2\left(F_{\mu \nu ; \beta \alpha} + F_{\mu \nu ;  \alpha \beta}\right)\dot{{\cal G}}^{\mu \beta}_{B11}{\cal G}^{\nu \alpha}_{B11} \nonumber\\
&&-\frac{iT^2}{24}F_{\lambda \nu}R^{\lambda}_{\, \, \, \alpha \beta\mu}\,\left(\dot{{\cal G}}^{\nu \mu}_{B11}\,{\cal G}^{\alpha \beta}_{B11}+\dot{{\cal G}}^{\alpha \mu}_{B11}\,{\cal G}^{\nu \beta}_{B11}+\dot{{\cal G}}^{\beta \mu}_{B11}\,{\cal G}^{\nu \alpha}_{B11}\right) \nonumber\\
&&+\frac{T}{12}R_{\mu \alpha \beta \nu}\left( \dot{{\cal G}}^{\mu \alpha}_{B11}\dot{{\cal G}}^{\beta \nu}_{B11}+\dot{{\cal G}}^{\mu \beta}_{B11}\dot{{\cal G}}^{\alpha\nu}_{B11}
+\Bigl(\ddot{{\cal G}}^{\mu \nu}_{B11}-2g^\mn \delta(0)\Bigr){\cal G}^{\alpha \beta}_{B11}
\right) 
\nonumber\\
&&-\frac{T^3}{6}F_{\alpha \beta;\gamma}F_{\mu \nu ;\delta}
\int^{1}_{0}d\tau_1\left(\dot{{\cal G}}^{\alpha \nu}_{B12}\, \dot{{\cal G}}^{\beta \mu}_{B12} \, {\cal G}^{\gamma \delta}_{B12}+\dot{{\cal G}}^{\alpha \nu}_{B12}\, {\cal G}^{\beta \delta}_{B12} \, \dot{{\cal G}}^{\gamma \mu}_{B12}\right) 
\Biggr\rbrace. \non\\
\label{resultscal}
\end{eqnarray}
Here $\bar\xi = \xi - \fourth$ where $\xi$ 
parametrizes the coupling of the loop scalar to the
scalar curvature (see appendix A for our conventions).
In the last term it is again understood that $\tau_2 = 0$.

\section{Effective Lagrangians at order $RF^4$}
\label{newresults}
\renewcommand{\theequation}{3.\arabic{equation}}
\setcounter{equation}{0}

To obtain the effective Lagrangians at a given order $O(F^n)$ from the
integral representations (\ref{resultspin}), (\ref{resultscal}), first one needs
to expand the worldline Green's functions to the required order. Adequate 
formulas for an arbitrary order have been given in appendix $B$ of \cite{badasc};
here in appendix \ref{green} we write down this expansion explicitly
to the order required for the present calculation. The integrals
are then elementary, and can be easily done using MATHEMATICA. However, the
form of the result is still highly redundant, and can be considerably reduced by an
application of the gauge and gravitational Bianchi identities. This is by far the most
laborious step of the procedure (we have found the program MathTensor very useful
for this task). We believe that the results given below are in the most compact form
which can be achieved by the use of these identities (further reduction may be possible
by the addition of total derivative terms, but we have not attempted this here).
We include also the order $O(F^2)$ terms for easy reference (although not the
pure Euler-Heisenberg terms).
Our conventions are given in appendix \ref{conventions}, where we
also collect some useful formulas.

\small{
\begin{eqnarray}
{\cal L}_{\rm scal}^{R(4)}&=&\frac{1}{16\, \pi^2}\frac{1}{m^2}\Bigg[ \frac{1}{12}\left(\bar\xi + \frac{1}{12}\right) R (F_{\mu \nu})^2 + \frac{1}{180}R_{\mu \nu} F^{\mu \alpha}F^{\nu}{}_\alpha \nonumber\\
&&-\frac{1}{72}R_{\mu \nu\alpha\beta}F^{\mu \nu} F^{\alpha \beta} - \frac{1}{180}(\nabla_{\alpha}F_{\mu \nu})^2 - \frac{1}{72}F_{\mu \nu}\square F^{\mu \nu} \Bigg] \nonumber\\
&&+\frac{1}{16\, \pi^2}\frac{1}{m^6}\Bigg[-\frac{1}{144}\left(\bar\xi +\frac{1}{12}\right)R(F_{\mu \nu})^4-\frac{1}{180}\left(\bar\xi +\frac{1}{12}\right)R\, \mbox{tr}[F^4] \nonumber\\
&&-\frac{1}{945 }R_{\alpha \beta}(F^4)^{\alpha \beta}+\frac{1}{1080}R_{\alpha \beta}(F^2)^{\alpha \beta}(F_{\gamma \delta})^2+\frac{1}{540}R_{\alpha \mu \beta \nu}(F^2)^{\alpha \beta}(F^2)^{\mu \nu} \nonumber\\
&&-\frac{1}{360}R_{\alpha \mu \beta \nu}(F^3)^{\alpha \mu}F^{\beta \nu}+\frac{1}{432}R_{\alpha \mu \beta \nu} F^{\alpha \mu} F^{\beta \nu}(F_{\gamma \delta})^2 \nonumber\\
&&-\frac{1}{540}(F^3)^{\mu \nu}\square F_{\mu \nu}+\frac{1}{432}F^{\mu \nu}\square F_{\mu \nu}(F_{\gamma \delta})^2-\frac{1}{1080}F_{\mu \nu ; \alpha \beta}(F^2)^{\alpha \beta}F^{\mu \nu}\nonumber\\
&&+\frac{1}{540}F_{\mu \nu ; \alpha \beta}(F^2)^{\alpha \nu}F^{\beta \mu}+\frac{1}{1080}(F_{\alpha \beta ; \gamma})^2 (F_{\mu \nu})^2 \nonumber\\
&&+\frac{1}{1890}F_{\alpha \beta ; \gamma}F_{\mu \nu ;}^{\ \ \ \gamma}F^{\alpha \mu}F^{\beta \nu}+\frac{1}{1890}F_{\alpha \beta ; \gamma}F_{\mu \ \ ; \delta}^{\ \alpha}F^{\beta \mu}F^{\gamma \delta} \nonumber\\
&&+\frac{2}{945}F_{\alpha \beta;}^{\, \ \ \ \mu}F_{\mu \  ; \delta}^{\  \alpha}(F^2)^{\beta \delta}-\frac{1}{1890}F_{\alpha \beta ; \gamma}F^{\ \beta ; \gamma}_{\mu}(F^2)^{\alpha \mu}\Bigg]\, , \nonumber\\
\nonumber\\
\label{LscalF4}
\end{eqnarray}
}

% L SPIN
\small{
\begin{eqnarray}
{\cal L}_{\rm spin}^{R(4)}&=&-\frac{1}{8\, \pi^2}\frac{1}{m^2}\Bigg[-\frac{1}{72} R (F_{\mu \nu})^2+\frac{1}{180}R_{\mu \nu} F^{\mu\alpha}F^{\nu}{}_\alpha \nonumber\\
&&+\frac{1}{36}R_{\mu\nu \alpha\beta}F^{\mu \nu} F^{\alpha \beta} - \frac{1}{180}(\nabla_{\alpha}F_{\mu \nu})^2 + \frac{1}{36}F_{\mu \nu}\square F^{\mu \nu} \Bigg] \nonumber\\
&& -\frac{1}{8\, \pi^2}\frac{1}{m^6}\Bigg[-\frac{1}{432}R(F_{\mu \nu})^4 +\frac{7}{1080}R\, \mbox{tr}[F^4]\nonumber\\
&&-\frac{1}{945 }R_{\alpha \beta}(F^4)^{\alpha \beta}-\frac{1}{540}R_{\alpha \beta}(F^2)^{\alpha \beta}(F_{\gamma \delta})^2+\frac{1}{540 }R_{\alpha \mu \beta \nu}(F^2)^{\alpha \beta}(F^2)^{\mu \nu} \nonumber\\
&&+\frac{11}{360}R_{\alpha \mu \beta \nu}(F^3)^{\alpha \mu}F^{\beta \nu}+\frac{1}{108}R_{\alpha \mu \beta \nu} F^{\alpha \mu} F^{\beta \nu}(F_{\gamma \delta})^2 \nonumber\\
&&-\frac{11}{945}F_{\alpha \beta ; \gamma}F^{\ \beta ; \gamma}_{\mu}(F^2)^{\alpha \mu}+\frac{2}{945}F_{\alpha \beta;}^{\, \ \ \ \mu}F_{\mu \  ; \delta}^{\  \alpha}(F^2)^{\beta \delta} \nonumber\\
&&+\frac{7}{270}(F^3)^{\mu \nu}\square F_{\mu \nu}+\frac{1}{108}F^{\mu \nu}\square F_{\mu \nu}(F_{\gamma \delta})^2 +\frac{1}{216}F_{\mu \nu ; \alpha \beta}(F^2)^{\alpha \beta}F^{\mu \nu} \nonumber\\
&&+\frac{1}{540}F_{\mu \nu ; \alpha \beta}(F^2)^{\alpha \nu}F^{\beta \mu}-\frac{1}{540}(F_{\alpha \beta ; \gamma})^2 (F_{\mu \nu})^2\nonumber\\
&&-\frac{2}{189}F_{\alpha \beta ; \gamma}F_{\mu \nu ;}^{\ \ \ \gamma}F^{\alpha \mu}F^{\beta \nu}-\frac{2}{189}F_{\alpha \beta ; \gamma}F_{\mu \ \ ; \delta}^{\ \alpha}F^{\beta \mu}F^{\gamma \delta} \Bigg].\nonumber\\
\nonumber\\
\label{LspinF4}
\end{eqnarray}
}

\section{Conclusions}
\label{conclusions}
\renewcommand{\theequation}{4.\arabic{equation}}
\setcounter{equation}{0}

To summarize, the effective Lagrangians (\ref{LscalF4}), (\ref{LspinF4}) constitute
the natural generalization of the Drummond-Hathrell Lagrangian (\ref{drumhath})
to the order $O(F^4)$ level, but still at linear order in the curvature, in Einstein-Maxwell
theory.
They contain the full information on the one -- loop amplitude involving 
four photons and one graviton, with a massive scalar or spinor in the loop,
in the limit where all photon and graviton energies are small compared
to the loop particle mass.  In future work, we hope to elaborate these amplitudes in
an explicit form, as a first step towards generalizing the $N$ -- photon 
amplitudes (\ref{resspin}) to the full $N$ -- photon/ $M$ -- graviton case.

\noindent
{\bf Acknowledgements:}
C.S. thanks S. Theisen and the Albert-Einstein Institute, Potsdam, for hospitality
during part of this work. We also thank G. Dunne for conversations and 
A. Avelino Huerta for computer help. 
J. M. D\'avila thanks CONACYT for financial support. 

\begin{appendix}

\section{Conventions and useful formulas}
\label{conventions}
\renewcommand{\theequation}{A.\arabic{equation}}
\setcounter{equation}{0}

In our conventions, the Einstein-Maxwell theory is described by  
\bea  
\Gamma[g,A] =    
\int d^D x\ \sqrt{g}\, \bigg (  
{1\over \kappa^2 } R - {1\over 4}F_{\mu\nu}F^{\mu\nu}  
\bigg )   
\label{EM}  
\eea  
where the metric $g_{\mu\nu}$ has signature $(-,+,+,\dots, +)$,   
$g= |{\rm det}\, g_{\mu\nu}|$, and $\kappa^2 = 16\pi G_N$.  
We use the following conventions for the curvature tensors,
\bear
[\nabla_\mu, \nabla_\nu] V^\lambda &=& 
R_{\mu\nu}{}^\lambda{}_\rho V^\rho \ , \ \ \ 
R_{\mu\nu}= R_{\lambda\mu}{}^\lambda{}_\nu 
\ , \ \  R= R^\mu{}_\mu > 0\ {\rm on\ spheres}\, , \non\\
\lbrack \nabla_\mu, \nabla_\nu \rbrack \phi &=& iF_{\mn} \phi  \, ,\non\\
\label{convcurv} 
\ear
where $V^{\mu}$ is an uncharged vector and $\phi$ a charged scalar.
The one-loop effective action for the scalar loop is defined by

\bear
\Gamma[g,A] &=& \ln {\Det}^{-1} (-\square_A +m^2 +\xi R)
\label{scaldet}
\ear
where $\square_A$ is the gauge and gravitational covariant laplacian for 
scalar fields. The parameter 
$\xi$ describes an additional non-minimal coupling to the scalar
curvature $R$. 
For the (Dirac) spinor loop, we define it by

\bear
\Gamma[g,A] 
&=&
\ln {\rm Det} (\rldd + m) 
\label{spindef}
\ear
where 

\bear
\rldd &=& \gamma^a e_a{}^\mu \nabla_{\!\mu} \ , \quad \quad 
\nabla_{\!\mu} = \partial_\mu + i e A_\mu 
+ {1\over 4}\omega_{\mu ab} \gamma^a \gamma^b 
\label{defcovder}
\ear
with $e_\mu{}^a$ the vielbein and
$ e= \det e_\mu{}^a$, $\omega_{\mu ab} $ the spin connection.

The following identities have been used for simplifying the effective
Lagrangians (\ref{LscalF4}),(\ref{LspinF4}):

\begin{equation}
F_{\mu \alpha; \beta}\,F^{\mu \beta; \alpha}=
\frac{1}{2}F_{\mu \beta ; \alpha}\,F^{\mu \beta ; \alpha} 
\, ,
\label{iddFdF}
\end{equation}

\begin{equation}
F^{\, \, \alpha}_{\mu}\,F^{\mu \beta}_{\, \, \ \	;\alpha \beta}=\frac{1}{2}F_{\mu \nu}\square F^{\mu \nu}
\, ,
\label{idFddF}
\end{equation}

\begin{equation}
F_{\mu \nu}\,F_{\alpha \beta}\,R^{\mu \alpha \nu \beta}=
\frac{1}{2}F_{\mu \nu}\,F_{\alpha \beta}\,R^{\mu \nu \alpha \beta}
\, ,
\end{equation}

\begin{equation}
(F^3)_{\mu \nu}\,F_{\alpha \beta}\,R^{\mu \alpha \nu \beta}=\frac{1}{2}(F^3)_{\mu \nu}\,F_{\alpha \beta}\,R^{\mu \nu \alpha \beta}
\, ,
\end{equation}

\begin{equation}
F_{\alpha \beta ; \mu}\,F^{\alpha \beta}_{\, \, \ \ ; \nu}\,(F^2)^{\mu \nu}=-2\,F^{\, \, \ \ \mu}_{\alpha \beta;}\,F^{\, \ \alpha}_{\mu \, \, \ ; \nu}\,(F^2)^{\beta \nu} 
\, ,
\end{equation}

\begin{equation}
F_{\alpha \beta ; \gamma}\,F^{\ \ \ \beta}_{\mu \nu ; }\,F^{\alpha \nu}\,F^{\gamma \mu}=-\frac{1}{2}F_{\alpha \beta ; \gamma}\,F^{\ \ \ \gamma}_{\mu \nu ;}\,F^{\alpha \mu}\,F^{\beta \nu}
\, ,
\end{equation}

\begin{equation}
F^{\ \ \ \mu}_{\alpha \beta ;}\,F_{\mu \nu ; \gamma}\,F^{\alpha \nu}\,F^{\beta \gamma}=-\frac{1}{2}F_{\alpha \beta ; \gamma}\,F^{\ \ \ \gamma}_{\mu \nu;}\,F^{\alpha \mu}\,F^{\beta \nu}
\, ,
\end{equation}

\begin{equation}
F^{\, \,\alpha}_{\mu}F^{\mu \beta}_{\, \, \ \ ;\alpha \beta}+F^{\, \,\alpha}_{\mu}F^{\mu \beta}_{\, \, \ \ ;\beta \alpha}=\frac{1}{2}F_{\mu \nu}F_{\alpha \beta}R^{\mu \nu \alpha \beta}+(F^2)^{\alpha \beta}R_{\alpha\beta}+F_{\mu \nu}\,\square F^{\mu \nu}
\, ,
\end{equation}

\begin{equation}
F^{\, \, \, \ \  \mu}_{\alpha \beta;}\,F^{\, \, \ \ \beta}_{\mu \nu;}(F^2)^{\alpha \nu}=-F^{\, \, \ \ \ \mu}_{\alpha \beta;}\,F^{\, \ \alpha}_{\mu \ \  ; \gamma}\,(F^2)^{\beta \gamma}-F_{\alpha \beta; \gamma}\,F^{\, \ \beta ; \gamma}_{\mu}\,(F^2)^{\alpha \mu}
\, ,
\end{equation}

\begin{equation}
F_{\alpha \beta ; \gamma}\,F^{\, \, \, \, \  \beta}_{\mu \nu ; }\,F^{\alpha \nu}\,F^{\gamma \mu}=F_{\alpha \beta ;\gamma}\,F^{\, \, \, \alpha}_{\mu \, \ ;\eta}\,F^{\beta \mu}\,F^{\gamma \eta}-F_{\alpha \beta ;\gamma}\,F^{\, \, \, \alpha}_{\mu \, \ ;\eta}\,F^{\beta \eta}\,F^{\gamma \mu}
\, ,
\end{equation}

\begin{equation}
F_{\alpha \beta ;\gamma}\,F^{\, \, \, \alpha}_{\mu \, \ ;\eta}\,F^{\beta \eta}\,F^{\gamma \mu}=\frac{1}{2}F_{\alpha \beta ; \gamma}\,F^{\ \ \ \gamma}_{\mu \nu ;}\,F^{\alpha \mu}\,F^{\beta \nu}+F_{\alpha \beta ; \gamma}\,F^{\ \alpha}_{\mu \ \, ; \eta}\,F^{\beta \mu}\,F^{\gamma \eta}
\, ,
\end{equation}

\begin{equation}
F^{\, \, \, \ \ \ \ \nu}_{\mu \nu ; \beta}\,F^{\beta \mu}+F^{\, \, \, \ \ \nu}_{\mu \nu ; \, \, \,\beta}\,F^{\beta \mu}=-\frac{1}{2}F_{\mu \nu}\,F_{\alpha \beta}\,R^{\mu \nu \alpha \beta}-(F^2)^{\alpha \beta}\,R_{\alpha \beta}-F_{\mu \nu}\,\square F^{\mu \nu}
\, ,
\end{equation}

\begin{equation}
(F^2)^{\alpha \nu}\,F^{\beta \mu}\,F_{\mu \nu ; \beta \alpha}-(F^2)^{\alpha \nu}\,F^{\beta \mu}\,F_{\mu \nu ; \alpha \beta}= \frac{1}{2}(F^3)^{\mu \nu}\,F^{\alpha \beta}\,R_{\mu \nu \alpha \beta}+(F^2)^{\mu \alpha}\,(F^2)^{\nu \beta}\,R_{\mu \nu \alpha \beta}
\, ,
\end{equation}

\begin{equation}
F^{\, \, \, \ \ \ \ \nu}_{\mu \nu ; \beta}\,(F^3)^{\beta \mu}+F^{\, \, \, \ \ \nu}_{\mu \nu ; \, \, \,\beta}\,(F^3)^{\beta \mu}=-\frac{1}{2}(F^3)_{\mu \nu}\,F_{\alpha \beta}\,R^{\mu \nu \alpha \beta}-(F^4)^{\alpha \beta}\,R_{\alpha \beta}-(F^3)_{\mu \nu}\,\square F^{\mu \nu} \label{idRFF}
\, .
\end{equation}

\no
The identities (\ref{iddFdF}) -- (\ref{idRFF}) are simple consequences of the Bianchi identities
\begin{eqnarray}
\nabla_{\alpha} F_{\beta\gamma} 
+ \nabla_{\beta} F_{\gamma\alpha} 
+ \nabla_{\gamma} F_{\alpha\beta} &=& 0 \, ,\label{bianchiF}\\
R_{\alpha\beta\gamma\delta} 
+ R_{\beta\gamma\alpha\delta} 
+R_{\gamma\alpha\beta\delta} &=& 0 \, . \label{bianchiR}
\end{eqnarray}

\section{Expansion of the field-dependent worldline Green's functions}
\label{green}
\renewcommand{\theequation}{B.\arabic{equation}}
\setcounter{equation}{0}

In this appendix we give the expansion of the constant field worldline Green's functions
${\cal G}_B,\dot{\cal G}_B,\ddot{\cal G}_B, {\cal G}_F, \dot {\cal G}_F$ 
to the order $O(F^4)$ required for the present computation. Defining

\bear
\bar G_{B12}  &:=& \abs{\tau_1 -\tau_2} - (\tau_1-\tau_2)^2
\label{defGbar}
\ear
those expansions can be written as

\begin{eqnarray}
\mathcal{G}_{B12}&=&\bar{G}_{B12}-\frac{1}{6}-\frac{i}{3}\dot{G}_{B12}\bar{G}_{B12}\mathcal{Z}+\left(\frac{1}{3}\bar{G}^2_{B12}-\frac{1}{90}\right)\mathcal{Z}^2 \nonumber\\&&
-\frac{i}{15}\bar{G}_{B12}\dot{G}_{B12}\left(\bar{G}_{B12}+\frac{1}{3}\right)\mathcal{Z}^3 
+\frac{1}{45}\left(2\bar{G}^2_{B12}\left(\bar{G}_{B12}+\frac{1}{2} \right) -\frac{1}{21}\right)\mathcal{Z}^4+\mathcal{O}(\mathcal{Z}^5) \, ,\nonumber\\&&
\end{eqnarray}

\begin{eqnarray}
\dot{\mathcal{G}}_{B12}&=&\dot{G}_{B12} + 2\,i\left(\bar{G}_{B12} - \frac{1}{6}\right)\mathcal{Z} + \frac{2}{3}\dot{G}_{B12}\bar{G}_{B12}\mathcal{Z}^2+i\left( \frac{2}{3}\bar{G}^2_{B12}-\frac{1}{45}\right)\mathcal{Z}^3 \nonumber\\
&&+\frac{2}{15}\bar{G}_{B12}\dot{G}_{B12}\left(\bar{G}_{B12}+\frac{1}{3} \right)\mathcal{Z}^4+\mathcal{O}(\mathcal{Z}^5) \, ,
\end{eqnarray}

\begin{eqnarray}
\ddot{\mathcal{G}}_{B12}&=&2\delta_{12}-2+2\,i\,\dot{G}_{B12}\mathcal{Z}-4\left(\bar{G}_{B12}-\frac{1}{6}\right)\mathcal{Z}^2+\frac{4}{3}i \bar{G}_{B12}\dot{G}_{B12}\mathcal{Z}^3\nonumber\\
&&-\left( \frac{4}{3}\bar{G}^2_{B12}-\frac{2}{45}\right)\mathcal{Z}^4+\mathcal{O}(\mathcal{Z}^5) \, ,
\end{eqnarray}

\begin{eqnarray}
\mathcal{G}_{F12}&=&G_{F12}-i\,G_{F12}\dot{G}_{B12}\mathcal{Z}+2G_{F12}\bar{G}_{B12}\mathcal{Z}^2-\frac{1}{3}i\,G_{F12}\dot{G}_{B12}\left(2\bar{G}_{B12}+1 \right)\mathcal{Z}^3 \nonumber\\
&&+\frac{2}{3}G_{F12}\bar{G}_{B12}\left(\bar{G}_{B12}+1\right)\mathcal{Z}^4+\mathcal{O}(\mathcal{Z}^5) \, ,
\end{eqnarray}

\begin{eqnarray}
\dot{\mathcal{G}}_{F12}&=&2\delta_{12}+2\,i\,G_{F12}\mathcal{Z}+2\,G_{F12}\dot{G}_{B12}\mathcal{Z}^2+4i\,G_{F12}\bar{G}_{B12}\mathcal{Z}^3\nonumber\\
&&+\frac{2}{3}G_{F12}\dot{G}_{B12}\left(2 \bar{G}_{B12}+1\right)\mathcal{Z}^4+\mathcal{O}(\mathcal{Z}^5) \, .
\end{eqnarray}

\end{appendix}

%%%%%%%%%%%%%%%%%%%%%%%%%%%%%%%%%%%%%%%%%%%%%%%%%%%%%%%%%%%%%%%%%%%%%%%%%%%%%%%

\end{document}